\documentclass[11pt]{article}

\usepackage[final]{acl}
\usepackage{times}
\usepackage{latexsym}

\usepackage[T1]{fontenc}

\usepackage[utf8]{inputenc}

\usepackage{microtype}

\usepackage{inconsolata}

\usepackage{graphicx}
\usepackage{makecell}
\usepackage{multirow}
\usepackage{amsmath,amssymb,amsfonts}
\usepackage{xurl}

\title{Cross-Lingual TTS for Indian Languages: A Benchmark and Inference-Time Duration Estimation Strategies}

\author{
  \textbf{Adarsh Arigala}\textsuperscript{*} \quad
  \textbf{Arjun Gangwar}\textsuperscript{*} \quad
  \textbf{Srinivasan Umesh} \\
  SPRING Lab, Indian Institute of Technology Madras, India \\
  \texttt{arigalaadarsh780@gmail.com} \quad
  \texttt{arjungangwar@gmail.com} \\
  \texttt{umeshs@ee.iitm.ac.in}
}

\begin{document}
\maketitle
\begingroup
\renewcommand{\thefootnote}{*}
\footnotetext{\textbf{equal contribution}}
\endgroup
\begin{abstract}
Recent alignment-free Text-to-Speech (TTS) models achieve high synthesis quality by removing explicit alignment modules and simplifying training. However, they often struggle in cross-lingual scenarios due to differences in character durations across languages, leading to duration mismatch and degraded speech quality. Existing approaches typically address this issue using explicit duration predictors, increasing model complexity. Systematic evaluation of TTS performance for Indian languages in cross-lingual settings remains under-explored due to the lack of standardized evaluation benchmarks. In this work, we construct a cross-lingual evaluation benchmark covering English and nine Indian languages and propose a language-aware inference-time duration estimation strategy for alignment-free TTS that incorporates language-specific duration-per character statistics without retraining or architectural modifications. We further develop heuristic formulations that refine duration estimation based on language-dependent duration variations. Experiments on bilingual and multilingual A-DMA models demonstrate improved speech naturalness, intelligibility, and speaker similarity while reducing duration mismatch. \footnote{The dataset and code recipe are available at:
\url{https://github.com/Cross-Lingual-TTS-for-Indian-Languages/Cross-Lingual-TTS-Indic-Test-Set/}
}

\end{abstract}

\section{Introduction}

Text-to-Speech (TTS) systems have achieved significant progress with the emergence of neural architectures, enabling natural and expressive speech synthesis. Recent alignment-free models such as E2-TTS~\cite{E2-TTS}, F5-TTS~\cite{chen-etal-2024-f5tts} and A-DMA~\cite{choi2025accelerating}, simplify training by removing explicit alignment modules while maintaining high synthesis quality. Although these models perform well when reference and target speech is in the same language, they often struggle in cross-lingual scenarios where reference and target languages differ. This happens due to differences in writing systems, orthographic characteristics, and language-specific duration-per character variations across languages. Moreover, in F5-TTS, duration estimation is based on the number of UTF-8 bytes obtained by encoding the input text, rather than the actual number of characters. Since characters from different writing systems occupy different numbers of UTF-8 bytes and languages exhibit different duration-per-character patterns, the estimated duration can become inaccurate, leading to duration mismatch and degraded synthesis quality.

To address these challenges, recent works have explored various strategies to improve duration modeling and generation quality in alignment-free TTS. Cross-Lingual F5-TTS~\cite{cross-lignual-f5-tts} removes the dependency on reference transcripts in flow-matching TTS and enables cross-lingual voice cloning through forced alignment during training. However, it introduces additional training complexity by requiring MMS-based alignment~\cite{MMS_Alignment} preprocessing and separately trained speaking-rate predictors, making the overall pipeline more complex than standard alignment-free TTS systems.

Another work, PFluxTTS~\cite{PFluxTTS} proposes a hybrid flow-matching framework that combines a duration-guided model and an alignment-free model through inference-time fusion, improving the trade-off between alignment stability and naturalness. However, it relies on an explicit duration predictor, which reduces the benefits of alignment-free TTS by reintroducing duration modeling.

Most people in India are bilingual and frequently switch between English and their regional languages during everyday conversations. Consequently, cross-lingual TTS between English and Indian languages has become increasingly important. These languages exhibit substantial differences in orthographic characteristics and character-to-speech duration, making accurate duration estimation essential for stable cross-lingual speech synthesis. Despite its practical importance, systematic evaluation of TTS performance for Indian languages in cross-lingual settings remains underexplored due to the lack of standardized evaluation benchmarks and limited investigation of duration adaptation strategies for alignment-free synthesis.
  
To address these limitations, we propose a cross-lingual evaluation benchmark covering nine Indian languages and English and introduce language-aware inference-time duration estimation strategies for alignment-free TTS. Our approach incorporates language-specific duration-per character statistics directly during inference without requiring retraining or additional model components. Specifically, we design heuristic formulations that adjust the estimated duration according to language-specific duration-per-character statistics, enabling effective cross-lingual duration adaptation.

This work makes the following contributions:

\begin{itemize}

\item We construct a cross-lingual evaluation benchmark  covering nine Indian languages and English, evaluating two settings: \textbf{(i) English Reference and Indic Generation}, where English reference speech is used to generate Indian language speech and \textbf{(ii) Indic Reference and English Generation}, where Indian language reference speech is used to generate English speech, to systematically analyze TTS performance in cross-lingual scenarios.

\item We train bilingual and multilingual A-DMA models for Indian languages from scratch to investigate their performance in cross-lingual settings.\footnote{\textbf{Demo Samples}: \url{https://cross-lingual-tts-for-indian-languages.github.io/Cross-Lingual-TTS-Demo/}}

\item We propose inference-time language-aware duration estimation strategies for alignment-free TTS in cross-lingual settings, addressing duration mismatch by incorporating language-specific duration-per character information into the generation process.

\end{itemize}
 

\begin{table}[!h]
\centering

\renewcommand{\arraystretch}{1.2}
\setlength{\tabcolsep}{2.0pt}
\small
\begin{tabular}{l c c c c }
\hline
\textbf{Language} & \textbf{Duration} & \textbf{No. of} & \textbf{No. of} & \textbf{Total}  \\
\textbf{Indic+Eng} & \textbf{(Hrs)} & \textbf{Indic Utt.} & \textbf{Eng Utt.} & \textbf{Utt.}  \\
\hline
Kannada    & 2.00 & 540 & 588 & 1128   \\
Telugu     & 2.00 & 531 & 588 & 1119 \\
Tamil      & 2.00 & 591 & 588 & 1179   \\
Hindi      & 1.35 & 171 & 588 & 759    \\
Malayalam  & 2.00 & 554 & 588 & 1142   \\
Punjabi    & 2.00 & 520 & 588 & 1108   \\
Bengali    & 2.00 & 543 & 588 & 1131   \\
Odia       & 2.00 & 532 & 588 & 1120   \\
Marathi    & 2.00 & 531 & 588 & 1119   \\
\hline
\end{tabular} 
\caption{Cross-lingual evaluation set statistics.}
\label{tab:cross_lingual_stats}
\end{table}

\section{Methodology}
\subsection{Baselines}
We adopt A-DMA~\cite{choi2025accelerating} framework for our experiments. It is based on the F5-TTS~\cite{chen-etal-2024-f5tts} backbone and introduces a training objective that combines conditional flow matching with dual-modality text and speech representation alignment losses to accelerate convergence. We utilize self-supervised representations (SSL) from the 21st layer of data2vec-AQC~\cite{data2vecAQC} to compute the speech-level alignment loss. The model is trained using the conditional flow matching loss from F5-TTS as the primary generative objective. A Connectionist Temporal Classification (CTC) based text alignment loss~\cite{CTC} is applied at an intermediate layer to for early character-level alignment, and a speech-level alignment loss is used at a later layer to enforce alignment in the speech representation space.
 \[
\mathcal{L} = 
\mathcal{L}_{\mathrm{CFM}}
+ \lambda_{\mathrm{text}} \mathcal{L}_{\mathrm{text}}
+ \lambda_{\mathrm{speech}} \mathcal{L}_{\mathrm{speech}},
\]
where $\lambda_{\mathrm{text}}=0.1$ and $\lambda_{\mathrm{speech}}=1.0$. This combined objective improves convergence and speech synthesis quality.

\subsection{Cross-lingual Evaluation Dataset}
\label{cross-lingual-dataset-preparation}
To test the cross-lingual generalization capability of the A-DMA~\cite{choi2025accelerating} model, we constructed a dedicated cross-lingual evaluation benchmark covering nine Indian languages and English. In this setting, the reference speech and the text being synthesized belong to different languages.

For each Indian-language and English pair in the cross-lingual evaluation setting, we constructed approximately two hours of evaluation data, allocating one hour to (i) English Reference and  Indic Generation and one hour to (ii) Indic Reference and English Generation. The Indian-language speech was selected from the Rasa~\cite{rasa} and IndicVoices-R~\cite{IndicVoices-R} test splits, while the English speech was sampled from the IndicTTS~\cite{indictts2023} training split. \footnote{Indian English is available only in IndicTTS-v2, so we take 500 utterances from its 250-hour training set as a held-out set. Each language has only two speakers, and all speakers also recorded Indian English.}

We considered two cross-lingual evaluation scenarios: (i) Indian-language reference speech with English generation text, and (ii) English reference speech with Indian-language generation text. In both scenarios, reference and target utterances were randomly sampled subject to duration constraints of 4–7 seconds and 4–10 seconds, respectively. Detailed statistics of the constructed benchmark, including the total number of utterances for each language pair, are provided in Table~\ref{tab:cross_lingual_stats}.



 \subsection{Model Setup}
 We trained the original A-DMA~\cite{choi2025accelerating} model for Indian languages in two settings: bilingual and multilingual. For the bilingual setup, each model was trained on one Indian language combined with a 30-hour subset of Indian English data~\cite{indictts2023}, while the multilingual model was trained on 24 languages as described in Section~\ref{trainig_set}.

The bilingual model used the F5-TTS Small architecture with 18 DiT layers~\cite{DiT}, 12 attention heads, and 768-dimensional embeddings, together with a 4-layer ConvNeXtV2~\cite{ConvNeXtV2} module with 512-dimensional embeddings, totaling 159M parameters. It was trained for 1.2M steps with 20k warmup updates and an effective batch size of 0.75 hours.

The multilingual model used F5-TTS Base with 22 DiT layers~\cite{DiT}, 16 attention heads, and 1024-dimensional embeddings, along with a 4-layer ConvNeXtV2~\cite{ConvNeXtV2} module with 512-dimensional embeddings, totaling 335.8M parameters. It was trained for 300k updates with 45k warmup updates and an effective batch size of 2.09 hours. For both settings, 100-channel mel spectrograms were extracted at a 24 kHz sampling rate with a hop size of 256 samples. Waveform synthesis was performed using the pre-trained Vocos vocoder~\cite{siuzdak2024vocos}, and optimization was carried out using AdamW~\cite{loshchilov2019decoupled} with a peak learning rate of $7.5 \times 10^{-5}$.
 
\begin{table}[!h]
\centering

\scriptsize
\renewcommand{\arraystretch}{1}
\setlength{\tabcolsep}{4pt}

\resizebox{\columnwidth}{!}{%
\begin{tabular}{lccccc}
\hline

\textbf{Language} &
\makecell{\textbf{Indic}\\\textbf{Voices-R}} &
\makecell{\textbf{Indic}\\\textbf{TTS-v2}} &
\makecell{\textbf{Indic}\\\textbf{TTS-v3}} &
\textbf{Rasa} &
\makecell{\textbf{Total}\\\textbf{(Hrs)}} \\

\hline

Assamese & 173.77 & 25.73 & 20.33 & 58.32 & 278.15\\
Bengali & 110.20 & 20.09 & 20.63 & 56.76 & 207.68\\
Bodo & 170.48 & 0 & 14.47 & 52.31 & 237.26\\
Gujarati & 4.50 & 26.49 & 18.34 & 25.39 & 74.71\\
Hindi & 72.75 & 26.23 & 22.51 & 0 & 121.49\\
Kannada & 43.22 & 22.32 & 11.54 & 54.62 & 131.70\\
Kashmiri & 63.07 & 0 & 0 & 0 & 63.07\\
Konkani & 49.33 & 0 & 11.00 & 26.33 & 86.66\\
Maithili & 80.32 & 0 & 9.73 & 29.34 & 119.39\\
Malayalam & 80.64 & 20.91 & 25.02 & 51.63 & 178.20\\
Manipuri & 21.00 & 17.65 & 16.97 & 0 & 55.62\\
Marathi & 48.73 & 22.37 & 22.78 & 55.36 & 149.24\\
Nepali & 104.31 & 0 & 7.24 & 55.09 & 166.64\\
Odia & 69.32 & 19.19 & 22.41 & 47.91 & 158.84\\
Punjabi & 71.62 & 11.21 & 22.63 & 55.84 & 161.30\\
Sanskrit & 33.25 & 0 & 22.31 & 51.47 & 107.03\\
Santali & 72.76 & 0 & 0 & 0 & 72.76\\
Sindhi & 9.03 & 0 & 14.53 & 0 & 23.56\\
Tamil & 98.03 & 26.59 & 23.14 & 29.94 & 177.70\\
Telugu & 134.86 & 36.71 & 23.67 & 52.27 & 247.50\\
Urdu & 76.36 & 0 & 0 & 0 & 76.36\\
Indian Eng. & 0 & 251.09 & 0 & 0 & 251.09\\
Dogri & 0 & 0 & 13.19 & 48.04 & 61.23\\
Rajasthani & 0 & 20.39 & 20.92 & 0 & 41.31\\

\hline
\textbf{Total}
& \textbf{1587.55}
& \textbf{546.97}
& \textbf{363.36}
& \textbf{750.62}
& \textbf{3248.50}\\

\hline
\end{tabular}
}
\caption{Training dataset statistics across multilingual speech corpora in hours.}
\label{tab:train_dataset_stats}

 \end{table}


\subsection{Language-Aware Heuristic Duration Estimation}

\subsubsection{Duration Estimation in \textbf{Original F5-TTS}}

Alignment-free TTS models require predicting target speech duration prior to synthesis. A-DMA uses the F5-TTS duration estimation formulation defined as:
\begin{equation}
D_{F5} = D_{ref} + \left\lfloor \frac{D_{ref}} {L_{ref}} \cdot \frac{L_{gen}}
{s}
\right\rfloor
\label{eq:f5_duration}
\end{equation}
where $D_{ref}$ denotes the reference speech duration, $L_{ref}$ and $L_{gen}$ represent the number of characters in reference and generation text, which are converted into UTF-8 byte encoding for duration estimation, and $s$ denotes the local speed factor, which we set to 1.

While the standard F5-TTS formulation performs well when reference and generated speech share the same language, it computes text length using UTF-8 byte-level encoding, which is not suitable for multilingual settings. In cross-lingual generation, this becomes problematic because Indian languages typically require multiple bytes per character, whereas English characters generally require a single byte. Consequently, byte-level length estimation leads to inconsistent scaling in duration prediction across languages. Previous works have addressed this limitation by incorporating explicit speaking-rate predictors~\cite{cross-lignual-f5-tts} or duration predictors~\cite{PFluxTTS}, which require modifications to the original F5-TTS architecture and model training.

To overcome this issue, we reformulate Original F5-TTS in a character-level space and report results using this variant, referred to as \textit{\textbf{Modified F5-TTS}}. Additionally, we propose two language-aware duration modeling strategies, \textit{Heuristic-1} and \textit{Heuristic-2}, which incorporate duration-per character information into the F5-TTS formulation through fusion and additive interaction mechanisms, respectively.



\subsubsection{\textbf{Heuristic-1}: Language-Aware Duration Model}

The heuristic (H1) extends the original F5 duration estimation by incorporating language-specific duration-per-character statistics.

\begin{equation}
\begin{aligned}
D_{H1} = D_{ref}
&+ \alpha \left\lfloor
\frac{D_{ref}}{L_{ref}} \cdot \frac{L_{gen}}{s}
\right\rfloor  \\
&+ \beta (D_{ind} + D_{eng})
\end{aligned}
\label{eq:h1}
\end{equation}
where $\alpha=0.7$ and $\beta=0.3$ in all experiments. The effect of varying $\alpha$ and $\beta$ is analyzed in Section~\ref{apha_beta_effect}. $D_{F5}$ denotes the duration estimated by F5-TTS. Duration estimations for Indian languages and English are computed as follows to capture language-dependent duration variations:


\begin{equation}
\begin{array}{c c}
D_{ind} = \dfrac{d_{ind}^{c}\times C_{ind}}{s}
\quad &
D_{eng} = \dfrac{d_{eng}^{c}\times C_{eng}}{s}
\end{array}
\label{eq:h1_decomp}
\end{equation}
 
where $D_{ref}$ denotes the reference speech duration,
$d_{ind}^{c}, d_{eng}^{c} $ denote the duration-per character statistics for Indic languages and English, respectively,
$C_{eng}$ and $C_{ind}$ denote the number of English and Indic characters in the generated text, and $s$ denotes the local speed factor, which we set to 1. 

The language-specific duration-per character are estimated from multilingual training corpora. For each utterance, the duration-per character is computed as:

\begin{equation}
d^{c} = \frac{ L_{audio} }{ C_{text} }
  = \frac{ N_{samples}/H }{ C_{text} }
\label{eq:duration-per-character}
\end{equation}

where $N_{samples}$ denotes the waveform length, $H$ denotes the hop length, and $C_{text}$ denotes the   text length in characters.

To obtain robust language-specific statistics, duration-per-character values are aggregated across all speakers within each language using the corresponding training corpus (Table~\ref{tab:train_dataset_stats}). For example, the Telugu statistic is computed from the Telugu training corpus, and similarly for the remaining languages. The language-specific average duration-per-character is computed as:

\begin{equation}
d^{c}_{lang} = \frac{1}{N_{\text{utt}}} \sum_{i=1}^{N_{\text{utt}}} D^{c}_i \label{eq:lang_avg}
\end{equation}
where $N_{\text{utt}}$ denotes the number of utterances per language. 

\begin{figure}[!h]
    \centering
    \includegraphics[width=\linewidth]{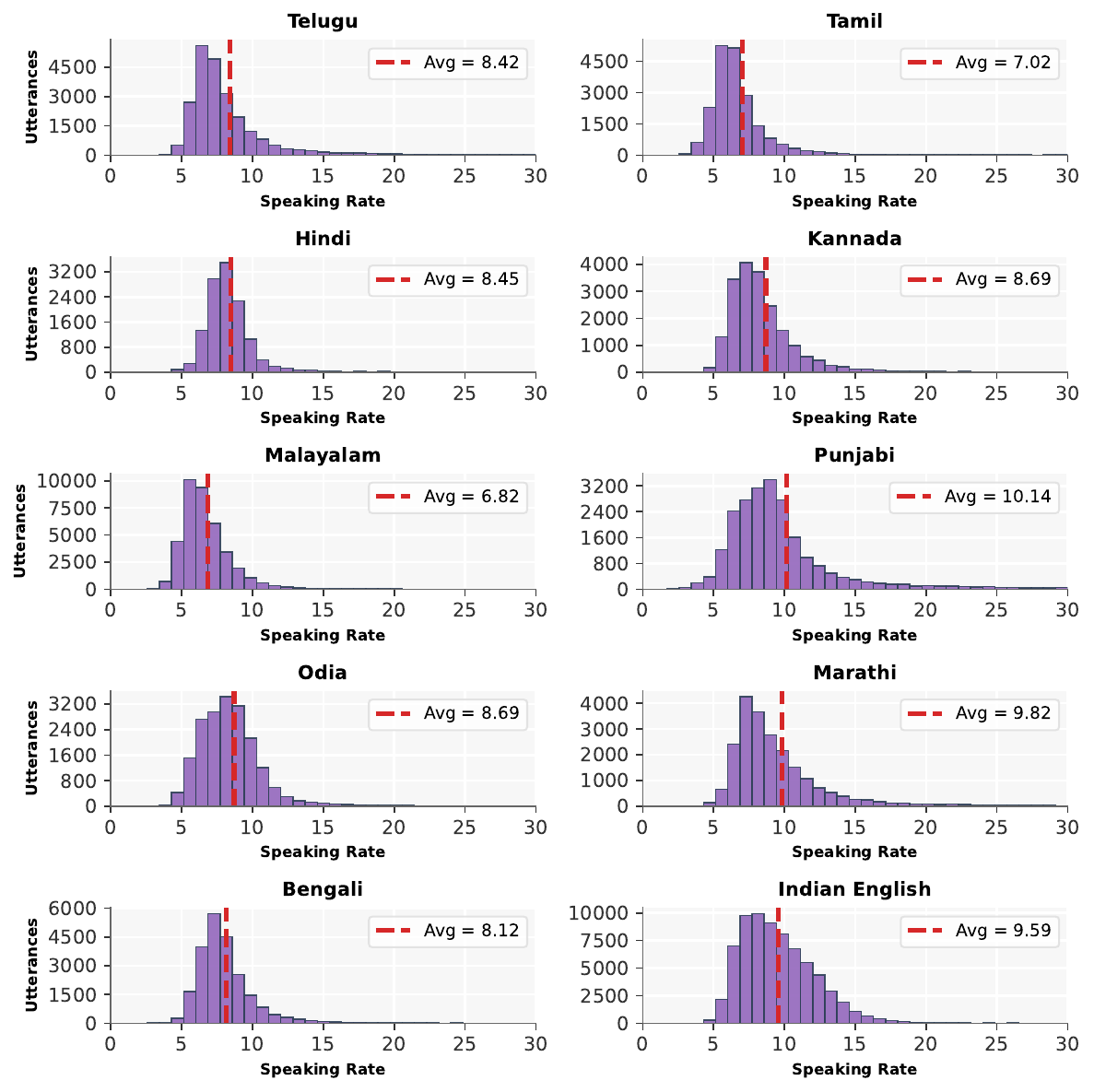}
    \caption{Utterance-level duration-per character distributions for the multilingual training data. The red dashed line indicates the average duration-per character ($d_{lang}^{c}$) used in the proposed heuristics.}
    \label{fig:sr_distribution}
   
\end{figure}

Fig.~\ref{fig:sr_distribution} shows the utterance-level distribution of duration-per-character values for all languages in the training corpora. Although each language exhibits natural variability due to differences in speakers and speaking styles, we use the average duration per character, $d_{lang}^{c}$, for each language in the proposed heuristics. The resulting language-specific average duration-per-character statistics are:

Indian English (9.6),
Kannada (8.7),
Telugu (8.4),
Tamil (7.0),
Hindi (8.5),
Malayalam (6.8),
Punjabi (10.1), 
Bengali (8.1),
Odia (8.7),
and Marathi (9.8). This formulation combines the original F5 duration estimate with language-aware duration estimation to better handle cross-lingual synthesis scenarios.

\begin{table*}[!t]
\centering

\renewcommand{\arraystretch}{0.9}
\setlength{\tabcolsep}{2pt}
\scriptsize

\resizebox{\textwidth}{!}{%
\begin{tabular}{l|cccc|cccc|cccc|cccc}
\hline

\multirow{2}{*}{\textbf{Model}}&
\multicolumn{4}{c|}{\textbf{Original F5-TTS}} &
\multicolumn{4}{c|}{\textbf{Modified F5-TTS }} &
\multicolumn{4}{c|}{\textbf{Heuristic-1  }} &
\multicolumn{4}{c}{\textbf{Heuristic-2  }} \\

\cline{2-17}

& WER & SIM & UTMOS & CER
& WER & SIM & UTMOS & CER
& WER & SIM & UTMOS & CER
& WER & SIM & UTMOS & CER \\

\hline

\multicolumn{17}{c}{\textbf{Kannada}}\\
\hline
GT
& 33.38 & -- & 3.461 & 5.88
& 33.38 & -- & 3.461 & 5.88
& 33.38 & -- & 3.461 & 5.88
& 33.38 & -- & 3.461 & 5.88 \\

 Bilingual
& 109.56 & 0.630 & 3.070 & 57.07
& 26.98 & 0.613 & 3.435 & 4.87
& 26.91 & 0.615 & 3.459 & 4.79
& 27.85 & 0.619 & 3.468 & 5.23 \\

 Multilingual
& 55.88 & 0.641 & 3.383 & 22.22
& 26.45 & 0.662 & 3.460 & 4.17
& 26.22 & 0.664 & 3.475 & 4.16
& 25.15 & 0.667 & 3.495 & 3.91 \\

\hline

\multicolumn{17}{c}{\textbf{Telugu}}\\
\hline
GT
& 21.06 & -- & 3.119 & 3.07
& 21.06 & -- & 3.119 & 3.07
& 21.06 & -- & 3.119 & 3.07
& 21.06 & -- & 3.119 & 3.07 \\

 Bilingual
& 82.05 & 0.662 & 3.249 & 48.02
& 20.68 & 0.653 & 3.575 & 3.49
& 21.15 & 0.656 & 3.590 & 3.66
& 23.00 & 0.658 & 3.592 & 5.33 \\

 Multilingual
& 37.66 & 0.635 & 3.436 & 18.02
& 16.73 & 0.664 & 3.508 & 2.08
& 16.40 & 0.662 & 3.508 & 1.93
& 16.16 & 0.668 & 3.554 & 1.96 \\

\hline

\multicolumn{17}{c}{\textbf{Tamil}}\\
\hline
GT
& 25.39 & -- & 3.357 & 4.87
& 25.39 & -- & 3.357 & 4.87
& 25.39 & -- & 3.357 & 4.87
& 25.39 & -- & 3.357 & 4.87 \\

 Bilingual
& 105.57 & 0.630 & 2.728 & 52.20
& 25.17 & 0.626 & 3.387 & 5.29
& 23.91 & 0.624 & 3.390 & 4.88
& 26.49 & 0.627 & 3.388 & 6.01 \\

 Multilingual
& 50.95 & 0.600 & 3.140 & 24.02
& 20.52 & 0.650 & 3.346 & 3.56
& 20.39	& 0.646	& 3.328 & 3.52
& 20.98 & 0.652 & 3.347 & 3.63 \\

\hline

\multicolumn{17}{c}{\textbf{Hindi}}\\
\hline
GT
& 14.40 & -- & 2.573 & 5.21
& 14.40 & -- & 2.573 & 5.21
& 14.40 & -- & 2.573 & 5.21
& 14.40 & -- & 2.573 & 5.21 \\

 Bilingual
& 62.45 & 0.633 & 3.101 & 47.01
& 12.58 & 0.662 & 3.608 & 4.04
& 13.16 & 0.656 & 3.627 & 4.30
& 12.87 & 0.664 & 3.638 & 4.31 \\

 Multilingual
& 26.70 & 0.666 & 3.435 & 17.05
& 11.80 & 0.679 & 3.555 & 3.61
& 11.66 & 0.680 & 3.566 & 3.42
& 11.71 & 0.685 & 3.585 & 3.49 \\

\hline

\multicolumn{17}{c}{\textbf{Malayalam}}\\
\hline
GT
& 25.89 & -- & 3.284 & 3.87
& 25.89 & -- & 3.284 & 3.87
& 25.89 & -- & 3.284 & 3.87
& 25.89 & -- & 3.284 & 3.87 \\

 Bilingual
& 121.85 & 0.579 & 2.301 & 56.82
& 28.91 & 0.622 & 3.359 & 6.22
& 26.49 & 0.622 & 3.403 & 5.03
& 31.42 & 0.624 & 3.377 & 7.73 \\

 Multilingual
& 53.42 & 0.600 & 3.170 & 23.81
& 22.98 & 0.661 & 3.343 & 3.49
& 22.15 & 0.660 & 3.351 & 3.34
& 22.45 & 0.667 & 3.353 & 3.44 \\

\hline

\multicolumn{17}{c}{\textbf{Punjabi}}\\
\hline
GT
& 15.75 & -- & 3.337 & 4.81
& 15.75 & -- & 3.337 & 4.81
& 15.75 & -- & 3.337 & 4.81
& 15.75 & -- & 3.337 & 4.81 \\

 Bilingual
& 71.81 & 0.639 & 3.265 & 51.17
& 16.14 & 0.632 & 3.607 & 4.08
& 17.12 & 0.639 & 3.620 & 5.69
& 17.29 & 0.641 & 3.615 & 6.18 \\

 Multilingual
& 31.65 & 0.674 & 3.475 & 17.62
& 15.53 & 0.681 & 3.592 & 4.44
& 15.49 & 0.687 & 3.623 & 4.35
& 15.38 & 0.688 & 3.574 & 4.45 \\

\hline

\multicolumn{17}{c}{\textbf{Bengali}}\\
\hline
GT
& 24.78 & -- & 3.062 & 6.48
& 24.78 & -- & 3.062 & 6.48
& 24.78 & -- & 3.062 & 6.48
& 24.78 & -- & 3.062 & 6.48 \\

 Bilingual
& 108.12 & 0.657 & 2.990 & 70.10
& 24.23 & 0.633 & 3.546 & 4.95
& 24.20 & 0.634 & 3.590 & 6.66
& 25.72 & 0.636 & 3.595 & 7.59 \\

 Multilingual
& 46.78 & 0.668 & 3.494 & 21.57
& 26.54 & 0.692 & 3.578 & 7.21
& 26.60 & 0.688 & 3.427 & 7.07
& 26.92 & 0.699 & 3.584 & 7.13 \\

\hline

\multicolumn{17}{c}{\textbf{Odia}}\\
\hline
GT
& 23.52 & -- & 3.131 & 4.51
& 23.52 & -- & 3.131 & 4.51
& 23.52 & -- & 3.131 & 4.51
& 23.52 & -- & 3.131 & 4.51 \\

 Bilingual
& 79.53 & 0.676 & 3.399 & 49.29
& 27.51 & 0.646 & 3.513 & 5.83
& 27.83 & 0.649 & 3.531 & 5.83
& 28.29 & 0.653 & 3.542 & 6.16 \\

 Multilingual
& 43.25 & 0.651 & 3.467 & 18.90
& 26.57 & 0.656 & 3.477 & 5.27
& 26.17	& 0.655	& 3.512	& 5.19
& 26.09 & 0.665 & 3.529 & 5.18 \\

\hline

\multicolumn{17}{c}{\textbf{Marathi}}\\
\hline
GT
& 10.56 & -- & 3.366 & 2.08
& 10.56 & -- & 3.366 & 2.08
& 10.56 & -- & 3.366 & 2.08
& 10.56 & -- & 3.366 & 2.08 \\

 Bilingual
& 74.51 & 0.596 & 2.900 & 43.51
& 11.46 & 0.605 & 3.475 & 2.47
& 11.72 & 0.611 & 3.496 & 2.47
& 12.04 & 0.605 & 3.480 & 2.76 \\

 Multilingual
& 30.76 & 0.637 & 3.449 & 15.44
& 10.58 & 0.657 & 3.501 & 2.08
& 10.74 & 0.661 & 3.528 & 2.00
& 10.43 & 0.663 & 3.545 & 1.96 \\

\hline

\end{tabular}
}
\caption{Cross-lingual TTS evaluation (English Reference and Indic Generation) across Indian languages}
\label{tab:cross_lingual_eng_to_ind}

\end{table*}


\subsubsection{\textbf{Heuristic-2}: Additive duration-per character Guided Scaling}

The second heuristic extends the original character-based duration estimation by incorporating relative duration-per character differences between languages through an additive correction mechanism. The target duration is computed as:
\begin{equation}
D_{H2} = D_{ref} +\left( \frac{D_{ref}}{C_{ref}} + \frac{d^{c}_{gen}}{d^{c}_{ref}} \right)  \frac{C_{gen}}{s} 
\label{eq:h2}
\end{equation}
 where $D_{ref}$ denotes the reference speech duration, $C_{ref}$ and $C_{gen}$ denote the number of characters in the reference and generated text, 
and $d^{c}_{ref}$ and $d^{c}_{gen}$ denote the language-specific duration-per character statistics corresponding to the reference and target generation languages. 
$\frac{D_{ref}}{C_{ref}}$ captures the reference duration-per-character, while $\frac{d^{c}_{gen}}{d^{c}_{ref}}$ provides a relative duration adjustment that introduces a language-dependent correction to per-character scaling across languages.







\section{Experimental Setup}

\subsection{Training Dataset}
\label{trainig_set}
 We follow the original A-DMA \cite{choi2025accelerating} training setup and adapt it for Indian languages in order to evaluate its multilingual performance. The model was trained using four multilingual speech corpora: IndicVoices-R~\cite{IndicVoices-R} containing 1,587.5 hours of speech from 21 languages; IndicTTS V2 and V3 versions~\cite{indictts2023} containing 547 hours and 363.3 hours of speech, respectively, covering 14 languages in V2 and 20 languages in V3; and Rasa~\cite{rasa} containing 750.6 hours of speech across 16 languages. In total, the model was trained on 24 languages, comprising approximately 3,248.5 hours of multilingual speech data. Detailed per-language training hours are provided in Table~\ref{tab:train_dataset_stats}. The trained model are evaluated using the proposed cross-lingual benchmark described in Section~\ref{cross-lingual-dataset-preparation}. 
 
\subsection{Evaluation Metrics} 
We evaluate synthesized speech using four objective metrics. Word Error Rate (WER) and Character Error Rate (CER) are computed from ASR transcriptions generated using Whisper-large-v3~\cite{radford2022whisper} for English and IndicConformer-600M ~\cite{indicconformer} for Indian languages. Speaker Similarity (SIM) evaluates speaker identity preservation using WavLM-large-based speaker embeddings~\cite{Chen2021WavLM}. UTMOS~\cite{UTMOS} measures speech naturalness via an objective MOS prediction model.
 
\begin{table*}[!t]
\centering

\renewcommand{\arraystretch}{0.9}
\setlength{\tabcolsep}{2pt}
\scriptsize

\resizebox{\textwidth}{!}{%
\begin{tabular}{l|cccc|cccc|cccc|cccc}
\hline

\multirow{2}{*}{\textbf{Model}} &
\multicolumn{4}{c|}{\textbf{Original F5-TTS }} &
\multicolumn{4}{c|}{\textbf{Modified F5-TTS }} &
\multicolumn{4}{c|}{\textbf{Heuristic-1 }} &
\multicolumn{4}{c}{\textbf{Heuristic-2 }} \\

\cline{2-17}

& WER & SIM & UTMOS & CER
& WER & SIM & UTMOS & CER
& WER & SIM & UTMOS & CER
& WER & SIM & UTMOS & CER \\

\hline

\multicolumn{17}{c}{\textbf{Kannada}}\\
\hline
GT
& 3.54 & -- & 4.248 & 1.33
& 3.54 & -- & 4.248 & 1.33
& 3.54 & -- & 4.248 & 1.33
& 3.54 & -- & 4.248 & 1.33 \\

 Bilingual
& 57.01 & 0.630 & 3.613 & 52.16
& 4.78 & 0.697 & 3.756 & 2.02
& 4.09 & 0.700 & 3.767 & 1.45
& 4.08 & 0.700 & 3.776 & 1.67 \\

 Multilingual
& 49.66 & 0.609 & 3.573 & 44.33
& 3.36 & 0.707 & 3.710 & 1.06
& 3.18 & 0.708 & 3.704 & 0.99
& 3.26 & 0.708 & 3.689 & 1.01 \\

\hline

\multicolumn{17}{c}{\textbf{Telugu}}\\
\hline
GT
& 3.54 & -- & 4.248 & 1.33
& 3.54 & -- & 4.248 & 1.33
& 3.54 & -- & 4.248 & 1.33
& 3.54 & -- & 4.248 & 1.33 \\

 Bilingual
& 52.09 & 0.626 & 3.181 & 47.81
& 3.86 & 0.722 & 3.414 & 1.53
& 3.60 & 0.720 & 3.374 & 1.24
& 3.74 & 0.724 & 3.408 & 1.45 \\

 Multilingual
& 45.64 & 0.628 & 3.254 & 41.46
& 3.65 & 0.721 & 3.289 & 1.20
& 3.73 & 0.724 & 3.293 & 1.18
& 3.50 & 0.725 & 3.257 & 1.25 \\

\hline

\multicolumn{17}{c}{\textbf{Tamil}}\\
\hline
GT
& 3.54 & -- & 4.248 & 1.33
& 3.54 & -- & 4.248 & 1.33
& 3.54 & -- & 4.248 & 1.33
& 3.54 & -- & 4.248 & 1.33 \\

 Bilingual
& 60.46 & 0.583 & 3.564 & 54.95
& 5.66 & 0.685 & 3.750 & 2.66
& 4.11 & 0.689 & 3.797 & 1.64
& 4.33 & 0.694 & 3.783 & 1.62 \\

  Multilingual
& 56.75 & 0.602 & 3.440 & 50.59
& 4.11 & 0.701 & 3.636 & 1.51
& 3.49 & 0.708& 3.635	&  1.14
& 3.35 & 0.706 & 3.635 & 1.14 \\

\hline

\multicolumn{17}{c}{\textbf{Hindi}}\\
\hline
GT
& 3.54 & -- & 4.248 & 1.33
& 3.54 & -- & 4.248 & 1.33
& 3.54 & -- & 4.248 & 1.33
& 3.54 & -- & 4.248 & 1.33 \\

 Bilingual
& 50.86 & 0.491 & 3.137 & 45.44
& 7.22 & 0.569 & 3.172 & 3.72
& 5.89 & 0.576 & 3.188 & 2.66
& 6.09 & 0.578 & 3.137 & 2.76 \\

 Multilingual
& 41.52 & 0.564 & 3.061 & 35.34
& 4.97 & 0.640 & 3.139 & 1.93
& 4.91 & 0.639 & 3.110 & 2.00
& 4.70 & 0.639 & 3.100 & 1.89 \\

\hline

\multicolumn{17}{c}{\textbf{Malayalam}}\\
\hline
GT
& 3.54 & -- & 4.248 & 1.33
& 3.54 & -- & 4.248 & 1.33
& 3.54 & -- & 4.248 & 1.33
& 3.54 & -- & 4.248 & 1.33 \\

 Bilingual
& 60.60 & 0.571 & 3.400 & 54.14
& 7.91 & 0.690 & 3.590 & 2.91
& 4.67 & 0.699 & 3.624 & 2.11
& 5.65 & 0.699 & 3.618 & 2.58 \\

 Multilingual
& 56.55 & 0.589 & 3.436 & 50.85
& 4.84 & 0.704 & 3.617 & 2.23
& 4.04 & 0.708 & 3.633 & 1.56
& 4.68 & 0.708 & 3.628 & 1.97 \\

\hline

\multicolumn{17}{c}{\textbf{Punjabi}}\\
\hline
GT
& 3.54 & -- & 4.248 & 1.33
& 3.54 & -- & 4.248 & 1.33
& 3.54 & -- & 4.248 & 1.33
& 3.54 & -- & 4.248 & 1.33 \\

 Bilingual
& 49.27 & 0.599 & 3.400 & 44.57
& 4.57 & 0.690 & 3.438 & 1.80
& 4.32 & 0.692 & 3.440 & 1.83
& 4.76 & 0.692 & 3.426 & 2.15 \\

 Multilingual
& 39.30 & 0.646 & 3.365 & 33.91
& 3.72 & 0.725 & 3.380 & 1.55
& 3.71 & 0.727 & 3.425 & 1.33
& 3.61 & 0.727 & 3.411 & 1.31 \\

\hline
\multicolumn{17}{c}{\textbf{Bengali}}\\
\hline
GT
& 3.54 & -- & 4.248 & 1.33
& 3.54 & -- & 4.248 & 1.33
& 3.54 & -- & 4.248 & 1.33
& 3.54 & -- & 4.248 & 1.33 \\

 Bilingual
& 52.98 & 0.571 & 3.407 & 48.31
& 4.19 & 0.661 & 3.444 & 1.68
& 4.15 & 0.663 & 3.407 & 1.81
& 4.45 & 0.663 & 3.462 & 2.15 \\

 Multilingual
& 47.22 & 0.595 & 3.324 & 42.60
& 4.14 & 0.687 & 3.325 & 1.54
& 4.19 & 0.687 & 3.315 & 1.51
& 4.06 & 0.687 & 3.282 & 1.42 \\

\hline

\multicolumn{17}{c}{\textbf{Odia}}\\
\hline
GT
& 3.54 & -- & 4.248 & 1.33
& 3.54 & -- & 4.248 & 1.33
& 3.54 & -- & 4.248 & 1.33
& 3.54 & -- & 4.248 & 1.33 \\

 Bilingual
& 50.99 & 0.599 & 3.151 & 45.84
& 5.29 & 0.687 & 3.277 & 2.32
& 5.15 & 0.689 & 3.288 & 2.17
& 5.23 & 0.699 & 3.287 & 2.39 \\

 Multilingual
& 46.57 & 0.606 & 3.243 & 41.51
& 4.17 & 0.691 & 3.317 & 1.50
& 3.93 & 0.708 & 3.634 & 1.44
& 3.92 & 0.695 & 3.314 & 1.44 \\

\hline

\multicolumn{17}{c}{\textbf{Marathi}}\\
\hline
GT
& 3.54 & -- & 4.248 & 1.33
& 3.54 & -- & 4.248 & 1.33
& 3.54 & -- & 4.248 & 1.33
& 3.54 & -- & 4.248 & 1.33 \\

 Bilingual
& 49.27 & 0.566 & 3.382 & 44.98
& 4.35 & 0.676 & 3.423 & 1.71
& 4.19 & 0.681 & 3.431 & 1.69
& 4.44 & 0.680 & 3.419 & 1.68 \\

 Multilingual
& 42.28 & 0.603 & 3.324 & 37.96
& 3.16 & 0.699 & 3.397 & 1.10
& 3.04 & 0.700 & 3.415 & 0.92
& 3.20 & 0.701 & 3.404 & 0.94 \\

\hline

\end{tabular}
}
\caption{Cross-lingual TTS evaluation (Indic Reference and English Generation) across Indian languages}
\label{tab:cross_lingual_ind_to_eng}

\end{table*}

 
\section{Results}

The trained bilingual and multilingual A-DMA models are evaluated using the proposed benchmark under four duration strategies: Original F5-TTS, Modified F5-TTS, Heuristic-1, and Heuristic-2. We evaluate performance in two cross-lingual settings: (i) Indic Reference and English Generation and (ii) English Reference and Indic Generation. Although recent methods such as Cross-Lingual F5-TTS~\cite{cross-lignual-f5-tts} and PFluxTTS~\cite{PFluxTTS} have proposed alternative duration modeling strategies, their official implementations were not publicly available at the time of our experiments. Consequently, we compare our proposed methods against the publicly available A-DMA baseline.

\subsection{\textbf{English Reference and Indic Generation}}Table~\ref{tab:cross_lingual_eng_to_ind} presents cross-lingual results for English reference and Indian-language speech generation.  The original F5-TTS exhibits severe degradation in cross-lingual settings, primarily due to byte-level text length estimation, which leads to incorrect duration scaling and high WER/CER across all languages. For instance, in Kannada, the bilingual model yields a WER/CER of 109.56/57.07, while the multilingual model also exhibits degraded performance, with a WER/CER of 55.88/22.22.

The Modified F5-TTS substantially reduces these errors across all languages by correcting text-length estimation, resulting in large relative improvements in both WER and CER. In Kannada, the multilingual model achieves relative reductions of 52.7\%/81.2\% in WER/CER compared to the Original F5-TTS, indicating better linguistic alignment. Further, Heuristic-1 and Heuristic-2 for the multilingual model, provide relative WER/CER reductions of 0.9\%/0.2\% and 4.9\%/6.2\% over the Modified F5-TTS, respectively, while UTMOS increases from 3.460 to 3.475 and 3.495. Similar trends are observed across all languages.



\subsection{\textbf{Indic Reference and English Generation}}Table~\ref{tab:cross_lingual_ind_to_eng} reports results for the reverse direction. The original F5-TTS again shows high error rates. For Kannada, the bilingual model yields a WER/CER of 57.01/52.16, while the multilingual model reduces these to 49.66/44.33, respectively.

The Modified F5-TTS significantly improves cross-lingual robustness across all languages. In Kannada, the multilingual and bilingual models achieves relative reductions of 93.2\%/97.6\% and 91.6\%/96.1\% in WER/CER over the Original F5-TTS. Further, Heuristic-1 and Heuristic-2 for the multilingual model, provide additional improvements over the Modified F5-TTS, achieving WER/CER reductions of 5.4\%/6.6\% and 3.0\%/4.7\%, respectively. SIM remains comparable, while UTMOS shows slight improvement. Similar trends are observed across all languages. Overall, all proposed duration strategies operate within a similar performance range, showing stable improvements across Indian languages. 

Compared to bilingual models, multilingual models generally achieve lower WER and CER across most languages, indicating improved robustness due to exposure to diverse scripts and duration-per character variations during training.

\begin{figure}[!h]  
    \centering
    \includegraphics[width=\linewidth  ]{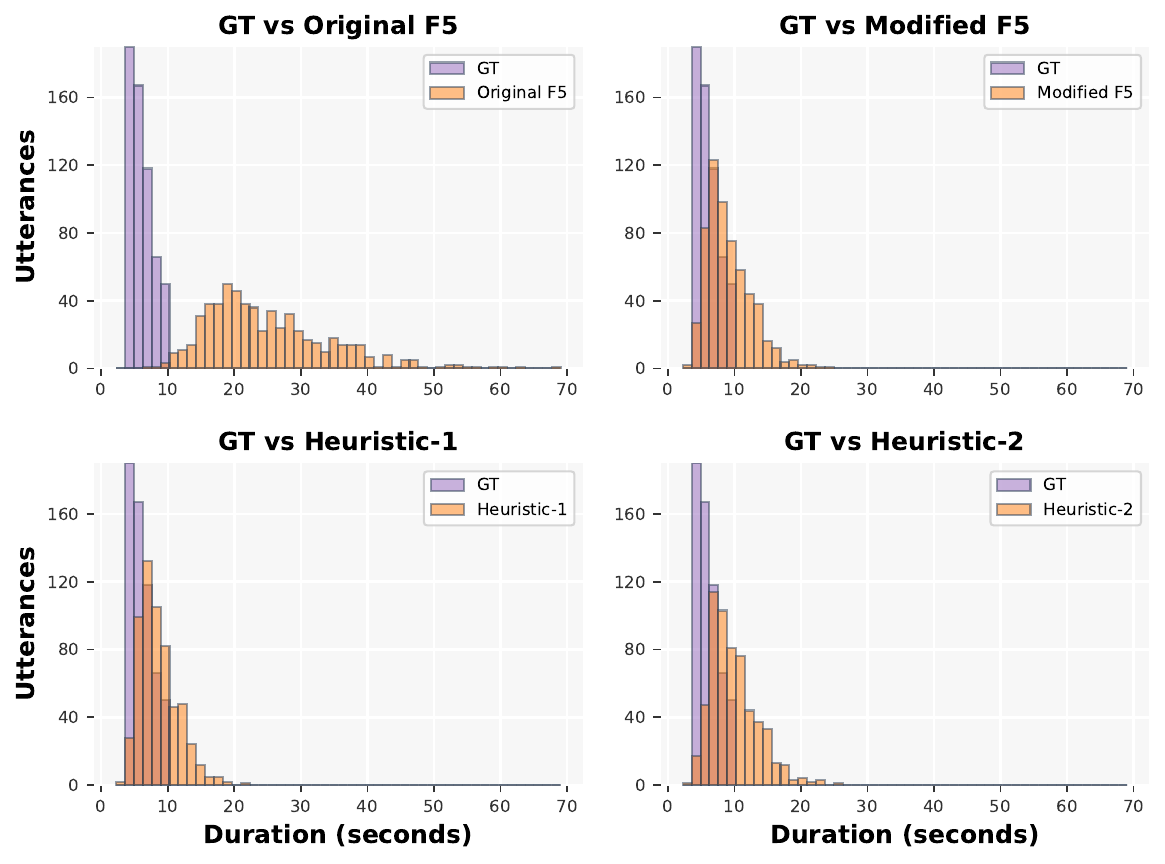}
    \captionsetup{font=small}
    \caption{Comparison of ground-truth (GT) and predicted durations for English reference and Tamil speech generation.}
    \label{fig:english_to_tamil_dur_analysis} 
\end{figure}

\section{Further Analysis}

\subsection{Qualitative Analysis}

\begin{figure}[!h]  
    \centering
    \includegraphics[width=\linewidth  ]{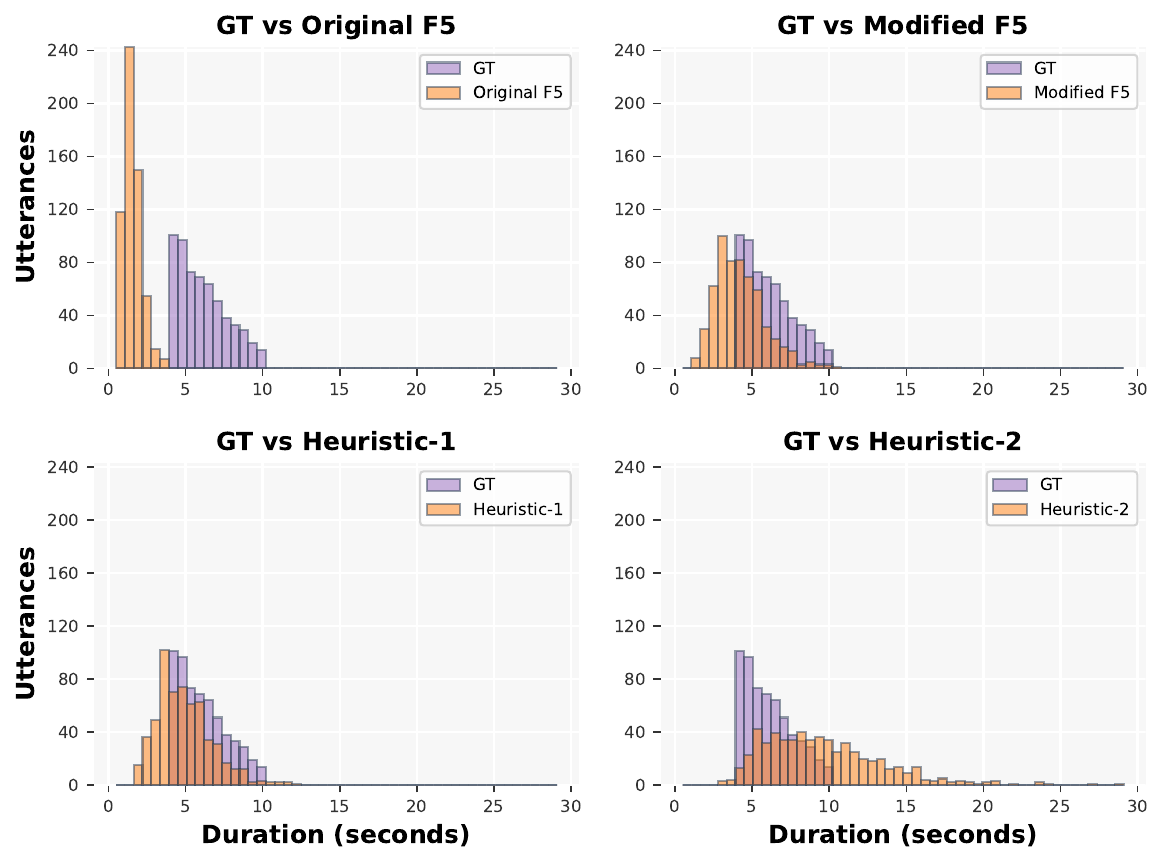}
    \captionsetup{font=small}
    \caption{Comparison of ground-truth (GT) and predicted durations for Tamil reference and English speech generation.}
    \label{fig:tamil_to_english_dur_analysis} 
    
\end{figure}

Fig.~\ref{fig:english_to_tamil_dur_analysis} and Fig.~\ref{fig:tamil_to_english_dur_analysis} show the predicted and ground-truth (GT) duration distributions on our benchmark for two cross-lingual settings: (i) English reference and Tamil speech generation and (ii) Tamil reference and English speech generation. GT durations correspond to natural multi-speaker utterances, while generated speech is conditioned on a reference utterance. Since GT and generated samples are not speaker-aligned, we compare duration distributions instead of absolute differences. The original F5-TTS  underestimates durations relative to GT due to byte-level UTF-8 encoding. The modified F5-TTS alleviates this issue by switching to character-level encoding; however, it still underestimates durations, resulting in speech that sounds too fast in setting (ii). Meanwhile, the proposed Heuristic-1 and Heuristic-2 methods produce durations that align better with the GT distribution in both settings. This indicates that incorporating language-specific duration-per character statistics during inference improves cross-lingual duration modeling and yields better alignment with ground-truth.

\begin{table*}[!h]
\centering

\renewcommand{\arraystretch}{0.8}
\setlength{\tabcolsep}{6pt}
\small

\begin{tabular}{l|cccc|cccc}
\hline

\multirow{2}{*}{\textbf{ $\alpha : \beta$}} 
& \multicolumn{4}{c|}{Ref-English and Gen-Indian}
& \multicolumn{4}{c}{Ref-Indian and Gen-English} \\
\cline{2-9}
& WER & SIM & UTMOS & CER & WER & SIM & UTMOS & CER \\

\hline
\multicolumn{9}{c}{Telugu} \\
\hline

GT     & 21.06 & -    & 3.119 & 3.07 & 3.54 & -    & 4.248 & 1.33 \\
6:4    & 16.75 & 0.661 & 3.243 & 1.98 & 3.21 & 0.725 & 3.379 & 1.03 \\
7:3    & 16.40 & 0.662 & 3.508 & 1.93 & 3.73 & 0.724 & 3.293 & 1.18 \\
8:2    & 16.43 & 0.662 & 3.243 & 2.05 & 3.49 & 0.723 & 3.379 & 1.21 \\
\hline
\multicolumn{9}{c}{Hindi} \\
  
\hline
GT     & 14.40 & -    & 2.573 & 5.21 & 3.54 & -    & 4.248 & 1.33 \\
6:4    & 11.72 & 0.678 & 3.608 & 3.48 & 4.41 & 0.642 & 3.172 & 1.71 \\
7:3    & 11.66 & 0.680 & 3.566 & 3.42 & 4.91 & 0.639 & 3.110 & 2.00 \\
8:2    & 11.70 & 0.676 & 3.608 & 3.46 & 4.79 & 0.642 & 3.172 & 1.75 \\
\hline
\multicolumn{9}{c }{Kannada} \\ 
 
\hline
GT     & 33.38 & -    & 3.461 & 5.88 & 3.54 & -    & 4.248 & 1.33 \\
6:4    & 25.70 & 0.660 & 3.435 & 4.00 & 3.20 & 0.708 & 3.756 & 0.97 \\
7:3    & 26.22 & 0.664 & 3.475 & 4.16 & 3.18 & 0.708 & 3.704 & 0.99 \\
8:2    & 26.04 & 0.660 & 3.435 & 4.12 & 3.19 & 0.709 & 3.756 & 1.00 \\
\hline
\end{tabular}
\caption{Heuristic-1 based cross-lingual TTS performance across $\alpha:\beta$ duration ratios.}
 
\label{tab:alpha_beta_results}
\end{table*}

\subsection{Effect of $\alpha:\beta$ ratio in Heuristic-1}
\label{apha_beta_effect}
To investigate the sensitivity of Heuristic-1 to the interpolation coefficients $\alpha$ and $\beta$, we evaluate three ratio configurations: 6:4, 7:3, and 8:2 using the multilingual model across Telugu, Hindi, and Kannada in both cross-lingual settings. As shown in Table~\ref{tab:alpha_beta_results}, the performance remains similar across ratio configurations, with minor improvements. For Telugu in the (i) English Reference and Indic Generation setting, the 7:3 ratio achieves the best WER/CER of 16.40/1.93, with relative improvements of 2.1\%/2.5\% over 6:4 and 0.2\%/5.9\% over 8:2, while SIM remains nearly unchanged.

In the reverse setting, (ii) Indic Reference and English Generation, the 6:4 ratio performs slightly better in some cases. Overall, the optimal ratio varies slightly with the language pair and direction, and we \textbf{recommend experimenting with different values}. We adopt $\alpha=0.7$ and $\beta=0.3$ for all experiments to ensure fair comparisons.


\section{Conclusion}
We propose a cross-lingual evaluation benchmark and language-aware duration strategies for alignment-free TTS. Our approach introduces Modified F5 for improved duration estimation and two language-aware duration prediction methods, Heuristic-1 and Heuristic-2, which incorporate language-specific duration-per character statistics. Extensive experiments on both bilingual and multilingual models, evaluated across nine Indian languages and two cross-lingual directions, show that the proposed heuristic methods consistently improve duration alignment. Qualitative analysis further highlights the importance of language-aware duration modeling for robust multilingual speech synthesis.

\section{Limitations}

Our approach has a few limitations that we plan to address in future work. The proposed duration estimation relies on language-level average duration-per character rates and modifies the inference process without retraining; therefore, it does not explicitly account for speaker- or utterance-level variations in speaking style and prosody. In addition, our evaluation covers nine Indian languages and English, focusing on English to Indic and Indic to English cross-lingual directions. As future work, we plan to extend the evaluation to Indic to Indic language pairs to investigate broader cross-lingual generalization. Finally, the proposed duration adjustments are heuristic rather than learned, and our comparison is primarily with the publicly available alignment-free TTS such as A-DMA/F5-TTS baselines, as implementations of some recent cross-lingual TTS methods were unavailable during our experiments.


\bibliography{custom}

@inproceedings{E2-TTS,
  author={Eskimez, Sefik Emre and Wang, Xiaofei and Thakker, Manthan and Li, Canrun and Tsai, Chung-Hsien and Xiao, Zhen and Yang, Hemin and Zhu, Zirun and Tang, Min and Tan, Xu and Liu, Yanqing and Zhao, Sheng and Kanda, Naoyuki},
  booktitle={IEEE Spoken Language Technology Workshop (SLT)}, 
  title={{E2 TTS: Embarrassingly Easy Fully Non-Autoregressive Zero-Shot TTS}}, 
  year={2024},
  volume={},
  number={},
  pages={682-689},
  doi={10.1109/SLT61566.2024.10832320}}

@inproceedings{chen-etal-2024-f5tts,
    author = "Chen, Yushen  and
      Niu, Zhikang  and
      Ma, Ziyang  and
      Deng, Keqi  and
      Wang, Chunhui  and
      JianZhao, JianZhao  and
      Yu, Kai  and
      Chen, Xie",
    title = {{F5-TTS: A Fairytaler that Fakes Fluent and Faithful Speech with Flow Matching}},
    booktitle = "Proc. 63rd Annual Meeting of the Association for Computational Linguistics (Volume 1: Long Papers)",
    year = "2025",
    address = "Vienna, Austria",
    publisher = "Association for Computational Linguistics",
    doi = "10.18653/v1/2025.acl-long.313",
    pages = "6255--6271",

}

@inproceedings{choi2025accelerating,
    author    = {Jeongsoo Choi and Zhikang Niu and Ji-Hoon Kim and Chunhui Wang and Joon Son Chung and Xie Chen},
    title     = {{Accelerating Diffusion-based Text-to-Speech Model Trainingwith Dual Modality Alignment}},
    year      = {2025},
    booktitle = {Proc. Interspeech},
    pages     = {3459--3463},
    doi       = {10.21437/Interspeech.2025-1236},
}

@inproceedings{cross-lignual-f5-tts,
  author={Liu, Qingyu and Chen, Yushen and Niu, Zhikang and Wang, Chunhui and Yang, Yunting and Zhang, Bowen and Zhao, Jian and Zhu, Pengcheng and Yu, Kai and Chen, Xie},
  booktitle={ IEEE International Conference on Acoustics, Speech and Signal Processing (ICASSP)}, 
  title={{Cross-Lingual F5-TTS: Towards Language-Agnostic Voice Cloning and Speech Synthesis}}, 
  year={2026},
  volume={},
  number={},
  pages={17362-17366},
  doi={10.1109/ICASSP55912.2026.11461636}}

@inproceedings{PFluxTTS,
  author={Pankov, Vikentii and Gribul, Artem and Tatanov, Oktai and Proskurov, Vladislav and Korotkova, Yuliya and Mylzenova, Darima and Vypirailenko, Dmitrii},
  booktitle={IEEE International Conference on Acoustics, Speech and Signal Processing (ICASSP)}, 
  title={{PFluxTTS: Hybrid Flow-Matching TTS with Robust Cross-Lingual Voice Cloning and Inference-Time Model Fusion}}, 
  year={2026},
  volume={},
  number={},
  pages={17612-17616}, 
  doi={10.1109/ICASSP55912.2026.11462483}}

@article{MMS_Alignment,
    author = {Pratap, Vineel and Tjandra, Andros and Shi, Bowen and Tomasello, Paden and Babu, Arun and Kundu, Sayani and Elkahky, Ali and Ni, Zhaoheng and Vyas, Apoorv and Fazel-Zarandi, Maryam and Baevski, Alexei and Adi, Yossi and Zhang, Xiaohui and Hsu, Wei-Ning and Conneau, Alexis and Auli, Michael},
    title = {{Scaling speech technology to 1,000+ languages}},
    year = {2024},
    issue_date = {January},
    publisher = {JMLR.org},
    volume = {25},
    number = {1},
    issn = {1532-4435},
    journal = {J. Mach. Learn. Res.},
    month = jan,
    articleno = {97},
    numpages = {52},
     
    }

@inproceedings{siuzdak2024vocos,
    author={Hubert Siuzdak},
    title={{Vocos: Closing the gap between time-domain and Fourier-based neural vocoders for high-quality audio synthesis}},
    booktitle={Proc. International Conference on Learning Representations (ICLR)},
    pages = {25719--25733},
    year={2024},
}

@inproceedings{radford2022whisper,
    author = {Radford, Alec and Kim, Jong Wook and Xu, Tao and Brockman, Greg and McLeavey, Christine and Sutskever, Ilya},
    title = {{Robust speech recognition via large-scale weak supervision}},
    booktitle = {Proc. International Conference on Machine Learning (ICML)},
    articleno = {1182},
    numpages = {27},
    pages = 	 {28492--28518},
    year = {2023},
}

@inproceedings{UTMOS,
    title     = {{UTMOS: UTokyo-SaruLab System for VoiceMOS Challenge}},
    author    = {Takaaki Saeki and Detai Xin and Wataru Nakata and Tomoki Koriyama and Shinnosuke Takamichi and Hiroshi Saruwatari},
    booktitle = {Proc. Interspeech},
    pages     = {4521--4525},
    doi       = {10.21437/Interspeech.2022-439},
    year      = {2022},
}

@article{Chen2021WavLM,
  author={Chen, Sanyuan and Wang, Chengyi and Chen, Zhengyang and Wu, Yu and Liu, Shujie and Chen, Zhuo and Li, Jinyu and Kanda, Naoyuki and Yoshioka, Takuya and Xiao, Xiong and Wu, Jian and Zhou, Long and Ren, Shuo and Qian, Yanmin and Qian, Yao and Wu, Jian and Zeng, Michael and Yu, Xiangzhan and Wei, Furu},
  journal={IEEE Journal of Selected Topics in Signal Processing, 2022}, 
  title={{WavLM: Large-Scale Self-Supervised Pre-Training for Full Stack Speech Processing}}, 
  volume={16},
  number={6},
  pages={1505-1518},
  doi={10.1109/JSTSP.2022.3188113},

  }

@inproceedings{ConvNeXtV2,
  author={Woo, Sanghyun and Debnath, Shoubhik and Hu, Ronghang and Chen, Xinlei and Liu, Zhuang and Kweon, In So and Xie, Saining},
  booktitle={ IEEE/CVF Conference on Computer Vision and Pattern Recognition (CVPR)}, 
  title={{ConvNeXt V2: Co-designing and Scaling ConvNets with Masked Autoencoders}}, 
  year={2023},
  pages={16133-16142},
  doi={10.1109/CVPR52729.2023.01548}}

@inproceedings{DiT,
  author={Peebles, William and Xie, Saining},
  booktitle={2023 IEEE/CVF International Conference on Computer Vision (ICCV)}, 
  title={Scalable Diffusion Models with Transformers}, 
  year={2023},
  volume={},
  number={},
  pages={4172-4182},
  doi={10.1109/ICCV51070.2023.00387}}

@inproceedings{loshchilov2019decoupled,
    author    = {Ilya Loshchilov and Frank Hutter},
    title     = {{Decoupled Weight Decay Regularization}},
    booktitle = {Proc. International Conference on Learning Representations (ICLR)},
    year      = {2019}
}

@inproceedings{rasa,
  title     = {{Rasa: Building Expressive Speech Synthesis Systems for Indian Languages in Low-resource Settings}},
  author    = {Praveen {Srinivasa Varadhan} and Ashwin Sankar and Giri Raju and Mitesh M Khapra},
  year      = {2024},
  booktitle = {Proc. Interspeech},
  pages     = {1830--1834},
  doi       = {10.21437/Interspeech.2024-2421},
  issn      = {2958-1796},
}

@article{IndicVoices-R,
  title={IndicVoices-R: Unlocking a Massive Multilingual Multi-speaker Speech Corpus for Scaling Indian TTS},
  author={Sankar, Ashwin and Anand, Srija and Varadhan, Praveen Srinivasa and Thomas, Sherry and Singal, Mehak and Kumar, Shridhar and Mehendale, Deovrat and Krishana, Aditi and Raju, Giri and Khapra, Mitesh},
  journal={NeurIPS 2024 Datasets and Benchmarks},
  year={2024}
}

@misc{indictts2023,
    title = {{Indic {TTS}: A Text-to-Speech Database for Indian Languages}},
    author = {Speech Technology Consortium and {Hema A Murthy} and {S Umesh}},
    year = {2023},
    publisher = {Indian Institute of Technology Madras},
    url = {https://www.iitm.ac.in/donlab/indictts/},
    institution = {Department of Computer Science and Engineering and Electrical Engineering, IIT MADRAS}
}

@inproceedings{data2vecAQC,
  author={Lodagala, Vasista Sai and Ghosh, Sreyan and Umesh, S.},
  booktitle={IEEE International Conference on Acoustics, Speech and Signal Processing (ICASSP)}, 
  title={{Data2vec-Aqc: Search for the Right Teaching Assistant in the Teacher-Student Training Setup}}, 
  year={2023},
  volume={},
  number={},
  pages={1-5},
  doi={10.1109/ICASSP49357.2023.10094726}}

@inproceedings{CTC,
    author = {Graves, Alex and Fern\'{a}ndez, Santiago and Gomez, Faustino and Schmidhuber, J\"{u}rgen},
    title = {{Connectionist Temporal Classification: labelling unsegmented sequence data with recurrent neural networks}},
    booktitle = { Proc. International Conference on Machine Learning (ICML)},
    doi = {10.1145/1143844.1143891},
    pages = {369–376},
    numpages = {8},
    year = {2006},
}

@misc{indicconformer,
  title = {{IndicConformerASR}},
  author = {{AI4Bharat}},
  year = {2024},
  url = {https://github.com/AI4Bharat/IndicConformerASR}
}

\end{document}